%% file: apogeekm_dwarf_arxiv.tex
\documentclass[fleqn,usenatbib]{mnras}
\usepackage{txfonts}

\usepackage[T1]{fontenc}
\usepackage{ae,aecompl}

\usepackage{graphicx}	
\usepackage{epstopdf}
\usepackage{pdflscape}

\newcommand{\nodata}{...}

\title[Colours of APOGEE K and M dwarfs]{Examining the relationships between colour, $T_{\rm eff}$, and [M/H] for APOGEE K and M dwarfs}

 \author[Schmidt, Wagoner, Johnson, et al.]{Sarah J. Schmidt$^{1,2}$, Erika L. Wagoner$^{1,3}$, Jennifer A. Johnson$^{1,4}$, 
 \newauthor James R. A. Davenport$^{5,6,7}$, Keivan G. Stassun$^{8,9}$, Diogo Souto$^{10}$, and Jian Ge$^{11}$ \\
$^1$ Department of Astronomy, Ohio State University, 140 West 18th Avenue, Columbus, OH 43210, USA\\
$^2$ Leibniz-Institute for Astrophysics Potsdam (AIP), An der Sternwarte 16, 14482, Potsdam, Germany\\
$^3$ Department of Physics, University of Arizona, 1118 E. Fourth Street, Tucson, AZ 85721,USA \\
$^4$ Center for Cosmology and Astro-Particle Physics, Ohio State University, Columbus, OH 43210, USA\\
$^5$ Department of Astronomy, University of Washington, Box 351580, Seattle, WA 98195 \\
$^6$ Department of Physics \& Astronomy, Western Washington University, Bellingham, WA 98225 \\
$^7$ NSF Astronomy and Astrophysics Postdoctoral Fellow \\
$^8$ Department of Physics and Astronomy, Vanderbilt University, Nashville, TN 37235, USA\\
$^9$ Department of Physics, Fisk University, Nashville, TN 37208, USA\\
$^{10}$ Observat\'ario Nacional, Rua General Jose Cristino, 77, 20921-400 S\~ao Crist\'ov\~ao, Rio de Janeiro, RJ, Brazil \\
$^{11}$ Astronomy Department, University of Florida, Gainesville, FL 32611, USA
}

\date{Accepted 11 May 2016}

\pubyear{2016}

\begin{document}
\label{firstpage}
\pagerange{\pageref{firstpage}--\pageref{lastpage}}
\maketitle

\begin{abstract}
We present the effective temperatures ($T_{\rm eff}$), metallicities, and colours in SDSS, 2MASS, and WISE filters, of a sample of 3834 late-K and early-M dwarfs selected from the Sloan Digital Sky Survey APOGEE spectroscopic survey ASPCAP catalog. We confirm that ASPCAP $T_{\rm eff}$ values between 3550~K$<T_{\rm eff}<$4200~K are accurate to $\sim$100~K compared to interferometric $T_{\rm eff}$ values. In that same $T_{\rm eff}$ range, ASPCAP metallicities are accurate to 0.18~dex between $-1.0<$[M/H]$<0.2$. For these cool dwarfs, nearly every colour is sensitive to both $T_{\rm eff}$ and metallicity. Notably, we find that $g-r$ is not a good indicator of metallicity for near-solar metallicity early-M dwarfs. We confirm that $J-K_S$ colour is strongly dependent on metallicity, and find that $W1-W2$ colour is a promising metallicity indicator. Comparison of the late-K and early-M dwarf colours, metallicities, and $T_{\rm eff}$ to those from three different model grids shows reasonable agreement in $r-z$ and $J-K_S$ colours, but poor agreement in $u-g$, $g-r$, and $W1-W2$. Comparison of the metallicities of the KM dwarf sample to those from previous colour-metallicity relations reveals a lack of consensus in photometric metallicity indicators for late-K and early-M dwarfs. We also present empirical relations for $T_{\rm eff}$ as a function of $r-z$ colour combined with either [M/H] or $W1-W2$ colour, and for [M/H] as a function of $r-z$ and $W1-W2$ colour. These relations yield $T_{\rm eff}$ to $\sim$100~K and [M/H] to $\sim$0.18~dex precision with colours alone, for $T_{\rm eff}$ in the range of 3550--4200~K and [M/H] in the range of $-$0.5--0.2.
\end{abstract}



\section{Introduction}
\label{sec:intro}
Late-K and M dwarfs are the most common stars in the Galaxy, dominating Galactic star counts at faint magnitudes. Because their 
lifetimes are longer than the age of the Universe, their numbers, compositions, positions, and motions provide a fossil record 
of the chemical and dynamical history of the Galaxy.  Upcoming large area photometric surveys will detect an unprecedented number of these low mass stars. It is critical to match photometric measurements of late-K and M dwarfs with their intrinsic properties so they can be used to understand Milky Way evolution. This includes, for example, an accurate and precise calibration of model isochrones so star formation histories can be correctly mapped into predicted star counts as a function of colour and apparent magnitude. The reliable fundamental properties of late-type stars are also of high importance for understanding the numerous planetary systems that have been identified around them; the mass and radius measurements for these planets are sensitive to uncertainties in the fundamental properties of their host stars. 

While equations relating photometric colours with properties such as T$_{\rm eff}$, metallicity,
and gravity have been determined for hotter stars in a number of filter systems \citep[e.g.,][]{Ramirez2005,GonzalezHernandez2009,Casagrande2010},
such correlations have been much more difficult to produce for the coolest dwarfs. 
Not only are K and M dwarfs fainter than solar-type stars, but the formation of molecules in their cool 
atmospheres results in complex optical spectra \citep[e.g.,][]{Valenti1998} that are challenging to model accurately. Recently, fundamental parameters have been determined for some of the brightest nearby late-K and M dwarfs from a combination of high S/N, high-resolution spectroscopy and interferometry \citep{Casagrande2008,Boyajian2012,Rajpurohit2013}. These parameters are not immediately useful for calibration, however, because they are frequently above the saturation limit for good survey photometry.

The most comprehensive relationships between photometry and fundamental properties have instead been based on low-resolution spectroscopy. \citet{Mann2015} used low-resolution infrared spectra to derive colour-$T_{\rm eff}$ relations for K7--M7 stars using spectrophotometrically derived $VR_CI_CgrizJHK_s$ and Gaia filters as part of their comprehensive work on bolometric corrections, radii, and masses for such stars.  \citet{Newton2014} also used low-resolution infrared spectroscopy to derive a photometric metallicity relation for M0--M7 using filters from the Two Micron All-Sky Survey \citep[2MASS;][]{skrutskie2006}. The relationship between colour and metallicity in SDSS $ugriz$ filters was most recently examined by \citet{Bochanski2013}, using metal-poor subdwarfs identified in SDSS low-resolution spectroscopy \citep{Savcheva2014}.

Therefore, the colours of cool dwarfs with measured $T_{\rm eff}$ and metallicity 
have not yet been determined observationally for the $ugriz$ filters and the mid-infrared filters recently used by the Wide-field Infrared Survey Explorer \citep[WISE;][]{wright2010}. Determining the association between the observed SDSS colours, including the $u$-band, and fundamental properties
of late-type stars is key to studying the stellar populations of SDSS, which, because of its relatively deep
photometry and large sky coverage off of the Galactic plane, is a rich source of M dwarfs \citep[e.g.,][]{Juric2008,Bochanski2010}. 

The high-resolution near-IR spectra of the Sloan Digital Sky Survey III \citep[SDSS-III][]{eisenstein11,dr10} Apache Point Observatory Galactic Evolution Experiment \citep[APOGEE;][]{majewski2015} observed late-type stars in $\sim 650$ fields. Each of these stars has photometry from 2MASS and WISE, and a subset of APOGEE stars were both located in the SDSS photometric footprint and faint enough that their SDSS $ugriz$ photometry is not saturated. We explore a limited range of late-K and early-M dwarfs with both APOGEE observations and high quality SDSS-2MASS-WISE photometry to relate the fundamental properties ($T_{\rm eff}$ and [M/H]) predicted by stellar population modeling with colours. This combination of colours and fundamental properties also provides important tests of current stellar isochrones. 

In Section~\ref{sec:apogee}, we describe the selection of our late-K and early-M (hereafter KM) dwarf sample and verify the $T_{\rm eff}$ and [M/H] from the APOGEE catalog. Section~\ref{sec:phot} discusses the SDSS, 2MASS, and WISE photometry and Section~\ref{sec:starmods} describes the model grids we use for comparison. In Section~\ref{sec:compare} we examine the relationships between colours, $T_{\rm eff}$, and [M/H] in both the models and the data and provide empirical relationships between $T_{\rm eff}$, [M/H] and colour. 

\section{APOGEE Spectroscopy}
\label{sec:apogee}
The APOGEE survey \citep{majewski2015}, part of SDSS-III \citep{eisenstein11,dr10},
uses a multi-object near-infrared spectrograph \citep{Wilson2010} operating on the
2.5-meter Sloan Foundation Telescope \citep{Gunn2006} at Apache Point Observatory.
The spectra cover most of the $H$-band from 1.51--1.70 $\mu$m with an average resolution of R$\sim$22500.
The targets were generally selected based on 2MASS $J-K_s$ dereddened colours and an $H$-band magnitude limit extending down to
13.8 mag, although most fields have $H\leq12.2$~mag \citep[see ][for a description of the target selection]{Zasowski2013}. Additional stars were specifically targeted for survey calibration. For the brightest stars, spectra were obtained by running a fiber from the NMSU 1-meter to the APOGEE instrument
and observing when the APOGEE instrument was not taking data at the 2.5-meter telescope \citep{Holtzman2015}. These
brighter stars were selected due to previously well determined properties, and include both Gaia benchmark stars \citep{Jofre2014} and 
stars with interferometric radii \citep{Boyajian2012}. As of SDSS-III Data Release 12 \citep[DR12;][]{Alam2015}, APOGEE obtained 618,080 spectra of 156,593 stars, primarily red giants used to trace Galactic structure. In general, the M dwarfs observed by APOGEE fell serendipitously into
the normal APOGEE colour and magnitude cuts as red stars, were targeted as SEGUE overlap targets, or were targeted as part of the M-dwarf ancillary project \citep{Deshpande2013}.

\subsection{APOGEE Spectroscopic Parameters}
\label{sec:verify}

Stellar parameters were measured from the $H$-band spectra by the APOGEE Stellar Parameters and Chemical Abundances Pipeline 
\citep[ASPCAP;][]{garcia2015}, which determines the $\chi^2$ minima between the observed spectra and a 6-dimensional grid of synthetic spectra \citep{Zamora2015}.
The six dimensions varied are T$_{\rm eff}$, log g, [M/H], [C/M], [N/M], and [$\alpha$/M] \citep[for additional detail see][]{Zamora2015}. The ranges spanned by the grid in 
DR12 are 3500K--8000K in T$_{\rm eff}$, 0 dex to 5 dex in log g, $-$2.5 dex to 0.5 dex in [M/H], and $-$1 to 1 for [C/M], [N/M],
and [$\alpha$/M]. The cool temperature edge is of particular concern for this work, as it limits our current effort
to late-K and early-M dwarfs. ASPCAP parameters become increasingly unreliable as the grid edge is approached; requiring reliable T$_{\rm eff}$ measurements essentially restricts our sample to stars with T$_{\rm eff}\geq3550$K. 

In the ASPCAP minimization, the [M/H] axis varies the abundances of all elements relative to the solar values. However, the [$\alpha$/M] axis independently varies the abundances of $\alpha$ elements (O, Mg, Si, S, Ca, Ti). Therefore, the [M/H] reported by ASPCAP is sensitive mainly to the
lines of iron-peak elements and maps well onto [Fe/H] values in the literature \citep{Meszaros2013,Holtzman2015}.

\input{apogeeKM_short_v4}

\citet{Holtzman2015} performs an extensive comparison of ASPCAP parameters for red giants with photometric T$_{\rm eff}$, seismic log g values, and literature metallicities for cluster stars. Due to the existence of these comparison values, the ASPCAP parameters for giants are calibrated to better match previous data. The ASPCAP values for dwarf stars were also examined, reveling systematically low $\log(g)$ values compared to isochrone predictions and difficulties properly treating rotation in the model grid. In Sections~\ref{sec:teff} and~\ref{sec:met} we compare the uncalibrated ASPCAP values for KM dwarfs to literature values to assess the reliability of ASPCAP parameters for studying the colours of these stars in additional filter sets.

\subsection{Selecting KM dwarfs from APOGEE Data Release 12}
\label{sec:samsel}
From the DR12 APOGEE catalog, we selected stars with effective temperatures of 3500~K~$\le T_{\rm eff} \le 4200$~K and $\log(g) \ge 4.0$. These stars overlap with spectral types K5--M2 \citep{Boyajian2012,Pecaut2013} and form the low temperature, low gravity edge of the ASPCAP grid; while there are cooler APOGEE targets, the DR12 release does not include their properties. We selected the best observations of stars that were targeted on multiple plates by excluding those with the EXTRATARG flag set to 4.

Our initial sample included 7784 stars, but we excluded 1139 spectra targeted as part of an ancillary program to examine embedded young cluster stars as those stars have peculiar colours. We then performed flag cuts to ensure the quality of the catalog parameters for the 6604 remaining stars. We excluded stars with $T_{\rm eff}$ and [M/H] flagged as bad, typically due to the proximity of a value to the edge of the model grid (most notably the lower T$_{\rm eff}$ boundary). We also excluded stars flagged for low S/N (corresponding to S/N per pixel $<70$), a warning or bad flag set due to possible rapid rotation, and high $\chi^2$ values (warn or bad). The sample of stars with reliable ASPCAP parameters includes 4246 stars; those with reliable extinction corrections (see Section~\ref{sec:ext}) are listed in Table~\ref{tab:all}\footnote{Table~\ref{tab:all} is also available via filtergtaph at \url{https://filtergraph.com/apogee_km_dwarf_colors}}.

\subsection{Accuracy of Effective Temperature}
\label{sec:teff}
\begin{figure}
\includegraphics[width=1.1\linewidth]{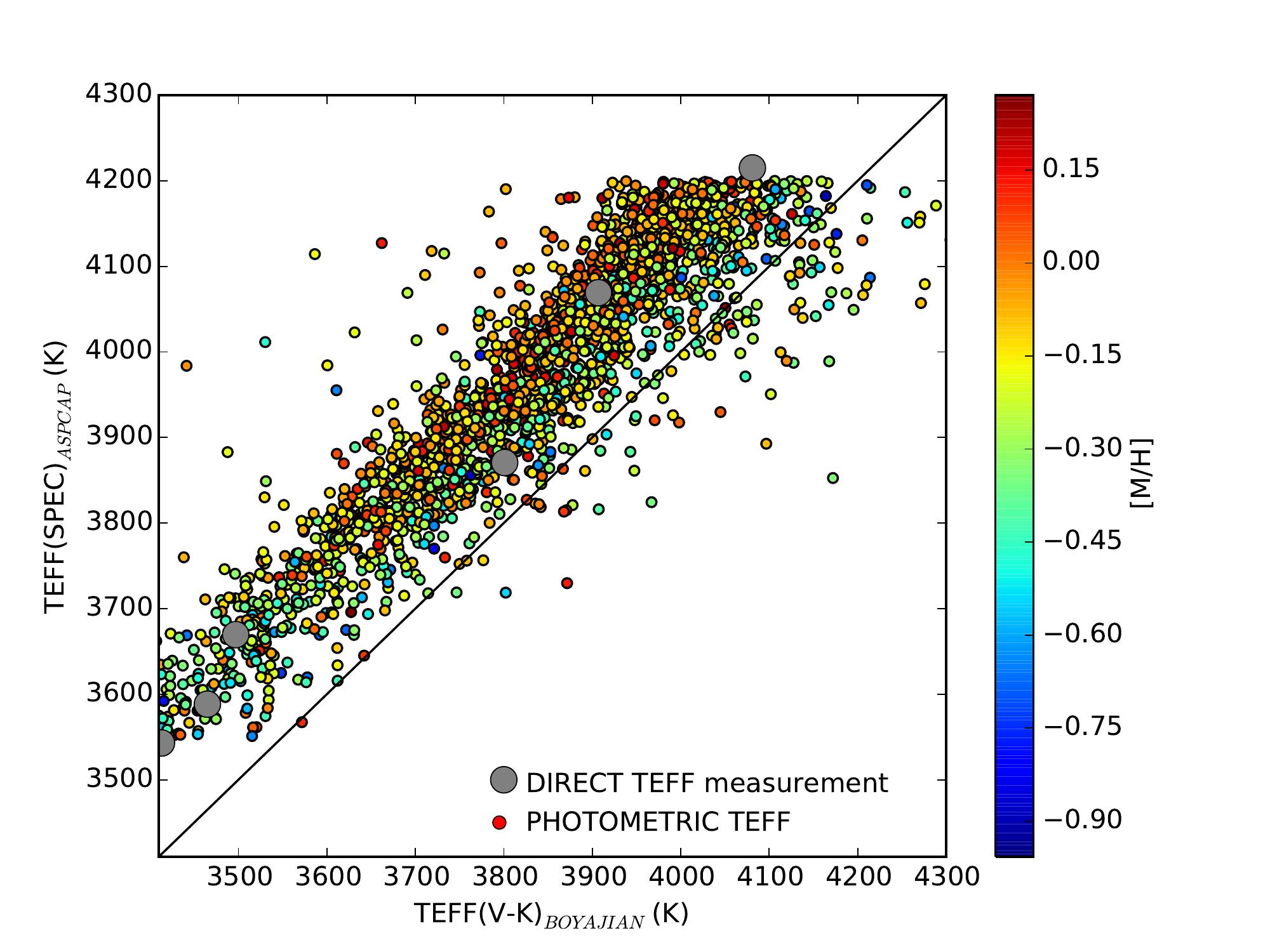}
\includegraphics[width=1.1\linewidth]{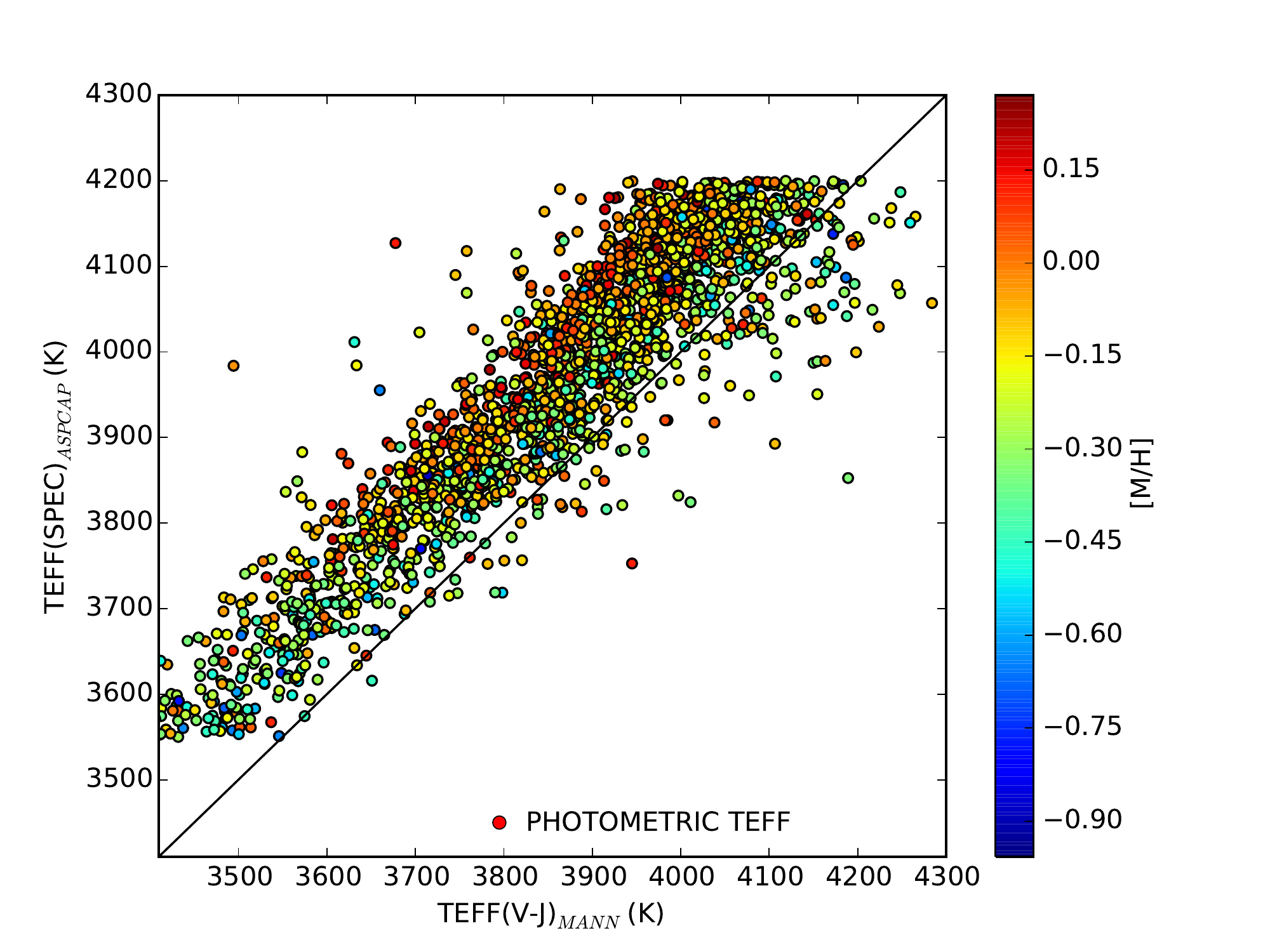}
\includegraphics[width=1.1\linewidth]{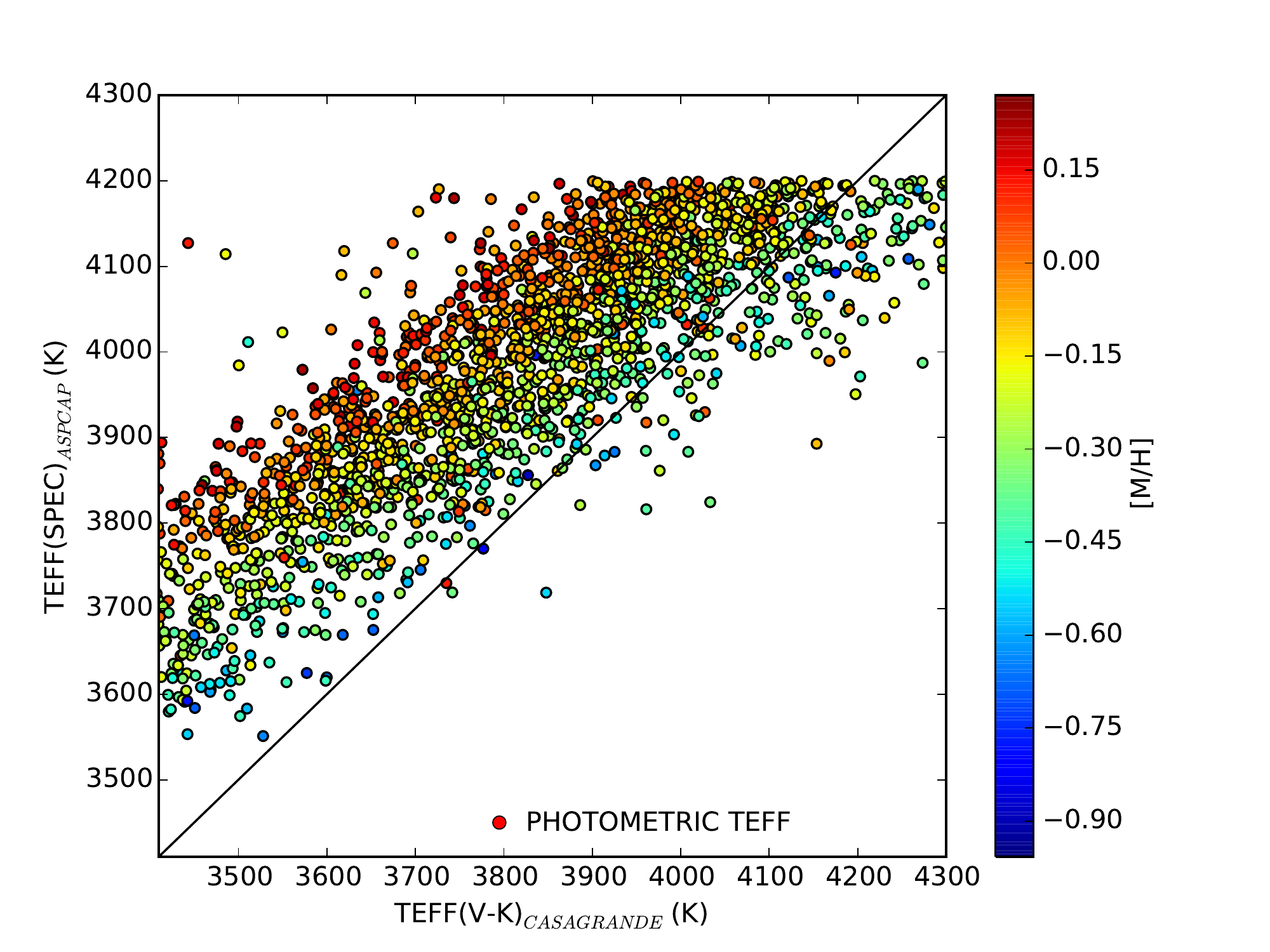}
\caption{ASPCAP T$_{\rm eff}$ compared to T$_{\rm eff}$ from multiple literature sources. In the top panel, the \citet[][]{Boyajian2013} interferometric (large grey circles) and $V-K$ (small coloured circles) T$_{\rm eff}$. In the middle panel, the \citep{Mann2015} colour-T$_{\rm eff}$ using $V-J$ colour  combined with ASPCAP [M/H] values (see their eqn. 6). The the bottom panel, the \citep{Casagrande2008} infrared flux technique T$_{\rm eff}$. In every panel, the points are colour-coded by ASPCAP [M/H] values and a one-to-one correspondence line is shown.}
\label{fig:teff_comp_all}
\end{figure}

We determined the accuracy of the ASPCAP $T_{\rm eff}$ values by comparing them to previously determined $T_{\rm eff}$ values derived from multiple sources, as shown in Figure~\ref{fig:teff_comp_all}. The most accurate and precise $T_{\rm eff}$ values are determined using a combination interferometric radii and bolometric luminosities. Five stars in the APOGEE KM sample (and one additional star with an ASPCAP $T_{\rm eff}=4215$~K) have interferometric $T_{\rm eff}$ measurements from \citet{Boyajian2013}; these stars are listed in Table~\ref{tab:compteff}, and their comparison is shown in the top panel of Figure~\ref{fig:teff_comp_all}. The ASPCAP $T_{\rm eff}$ values are $130$~K hotter than the interferometric $T_{\rm eff}$ values, with an rms scatter of 30~K. The comparison between these six values and the ASPCAP values reveals no systematic dependence on [M/H] or T$_{\rm eff}$, but the sample size is too small to rule out systematic issues. 

\begin{table}
\centering
\caption{Comparison of ASPCAP T$_{\rm eff}$ with T$_{\rm eff}$ from Interferometric Radii } 
\label{tab:compteff} 
\begin{tabular}{lllrr}
\hline
2MASS ID & Other & Interfer.$^{\rm a}$  & \multicolumn{2}{c}{ASPCAP} \\ 
       &        Name        & $T_{\rm eff}$ (K) & $T_{\rm eff}$ (K)  & [M/H] \\
         \hline
05312734$-$0340356 & GJ205 & 3801 $\pm$ 9 & 3871 & 0.16 \\ 
09142298+5241125 & GJ338A & 3907 $\pm$ 35 & 4069 & $-$0.12\\
10112218+4927153 & GJ380 & 4081 $\pm$ 15 & 4215 & 0.02\\
11032023+3558117 & GJ411 &3465 $\pm$ 17 & 3588 & $-$0.71\\
11052903+4331357 & GJ412A & 3497 $\pm$ 39 & 3670 & $-$0.60\\
17362594+6820220 & GJ687 & 3413 $\pm$ 28 & 3543 & $-$0.08\\
\hline
\multicolumn{5}{l}{$^{\rm a}$ From \citet{Boyajian2013}}
\end{tabular}
\end{table}

In addition to providing individual measurements, \citet{Boyajian2013} also used interferometric radii and bolometric luminosities to calibrate a $V-K$/$T_{\rm eff}$ relation. We combined $V$-band photometry for 2446 APOGEE stars in our sample from the AAVSO Photometric All Sky Survey \citep[APASS;][]{Henden2014} Data Release 8 (corrected for extinction as described in Section~\ref{sec:ext}) with 2MASS $K_S$ magnitudes (see Section~\ref{sec:phot}) to derive photometric $T_{\rm eff}$ values based on that relation; the comparison is shown in the top panel of Figure~\ref{fig:teff_comp_all} and the $V$ magnitudes and calculated $T_{\rm eff}$ values are included in Table~\ref{tab:all}. The formal uncertainties on the $V-K$ $T_{\rm eff}$ values are small ($\sim$30~K) due to low photometric uncertainties (0.02--0.03~mag) and a small scatter in the relation \citep[2\%;][]{Boyajian2013}. Overall, the ASPCAP $T_{\rm eff}$ values are 130~K hotter than the photometric $T_{\rm eff}$ values, with an rms scatter of 81~K. There is no evidence of a metallicity dependence in the comparison between the two sets of values. 

\citet{Mann2015} derived colour-$T_{\rm eff}$ relations for M dwarf using the method described by \citet{Mann2013a} that relies on the comparison of low-resolution infrared spectra to the BT-Settl version of the PHOENIX atmosphere models \citep{allard03}. We compared the ASPCAP $T_{\rm eff}$ values to those calculated from the \citet{Mann2015} relation as a function of $V-J$ colour that includes an explicit [M/H] term (shown in the middle panel of Figure~\ref{fig:teff_comp_all} and included in Table~\ref{tab:all}). The ASPCAP values are 101~K hotter with an rms scatter of 79~K\footnote{We note that the agreement is significantly poorer with the \citet{Mann2015} relation that uses $J-H$ as a metallicity proxy, likely because there is a degeneracy between $T_{\rm eff}$ and [M/H] in the $V-J$/$J-H$ plane.} The offset and scatter are similar to the \citet{Boyajian2013} offset because the \citet{Mann2013a} method for determining $T_{\rm eff}$ values was explicitly tuned to best match the \citet{Boyajian2013} relations. 

\citet{Casagrande2008} calculate $T_{\rm eff}$ values based on the infrared flux technique and derive a relationship based on $V-K$ colour. The bottom panel of Figure~\ref{fig:teff_comp_all} shows these $T_{\rm eff}$ values compared to ASPCAP $T_{\rm eff}$ values. The mean agreement is poor; the ASPCAP values are 161~K warmer than those from the $V-K$/$T_{\rm eff}$ relation and have a scatter of 140~K. The lack of agreement stems primarily from the absence of [M/H] from the $T_{\rm eff}$ calculation; higher metallicity stars ([M/H]$\sim0$) have ASPCAP $T_{\rm eff}$ values that are 200--300~K hotter their $V-K$ $T_{\rm eff}$ values, while lower metallicity stars ([M/H]$\sim-0.6$) fall closer to the 1:1 line.

Based on these comparisons, it is clear that there is a metallicity dependance that must be taken into account to calculate accurate $T_{\rm eff}$ values using photometry. From the comparison with the \citet{Boyajian2013} values, it is likely that the ASPCAP $T_{\rm eff}$ values for KM dwarfs are overestimated by $\sim$130~K. We discuss the effect of this offset as part of Section~\ref{sec:compare}. 

We used 25 APOGEE KM stars with duplicate observations that pass our quality cuts as an additional check on the uncertainty. The duplicate observations were on average 22~K warmer with a dispersion of 74~K. We adopt an overall $T_{\rm eff}$ uncertainty of 100~K based on both the duplicate observations and the scatter in the comparison between the APOGEE KM values and those from both \citet{Boyajian2013} and \citet{Mann2015}. 

\subsection{Accuracy of Metallicities}
\label{sec:met}

\begin{figure}
\includegraphics[width=1.1\linewidth]{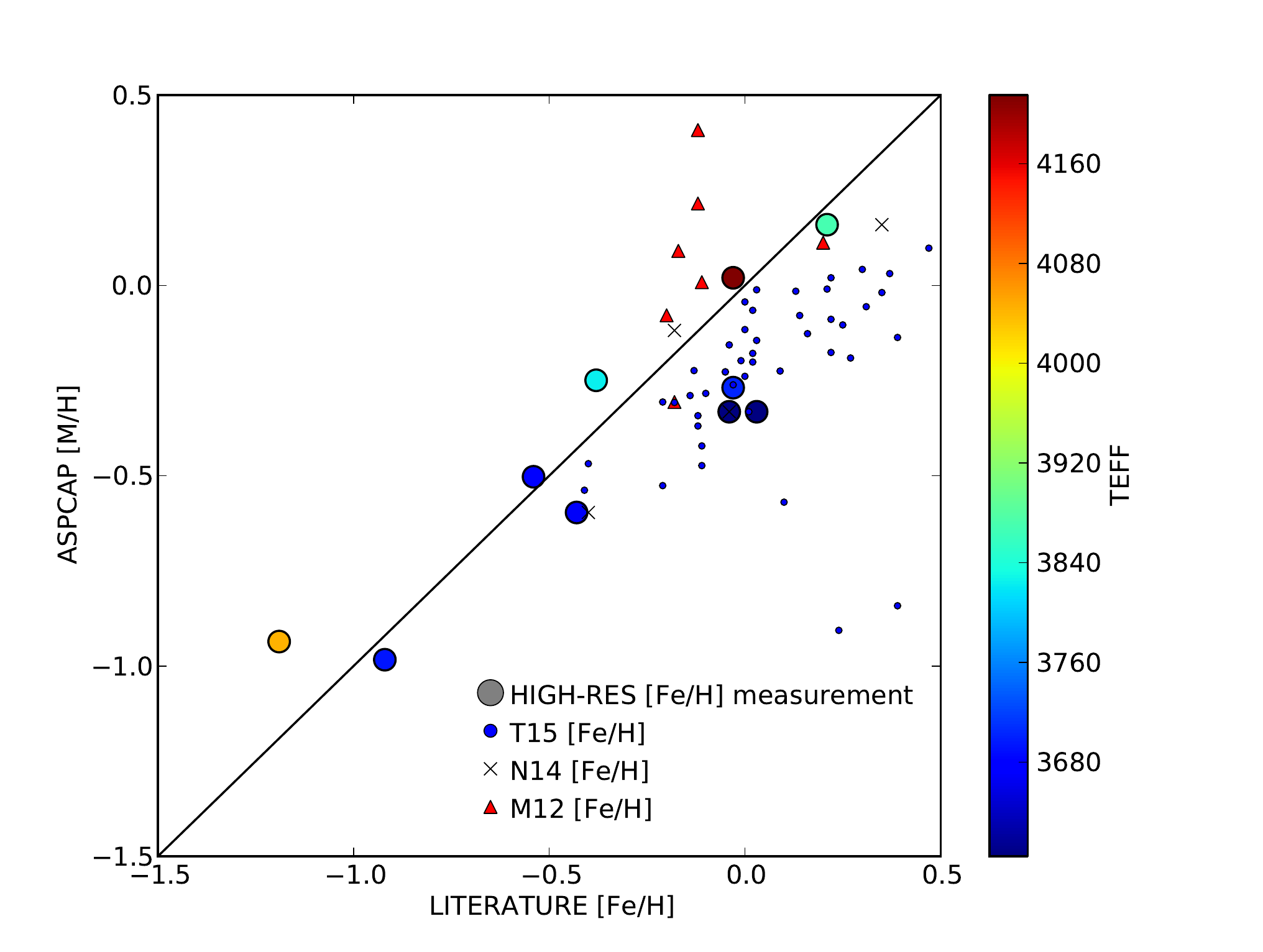}
\caption{ASPCAP [M/H] values compared to literature [Fe/H] values for high-resolution literature sample (filled circle colour-coded by T$_{\rm eff}$). We also show the comparison for stars we have in common with the measurements
based on the calibration of low-resolution spectra from \citet{Terrien2015}, \citet{Newton2014}, and \citet{Muirhead2012}. The overall comparison with other high-resolution analyses is good.}
\label{fig:metcomp}
\end{figure}

To test the accuracy of the ASPCAP [M/H] values, we compared them to measurements of the 
metallicities of M dwarfs derived from either high-resolution spectroscopic
analysis of individual M dwarfs or of hotter primaries in binary systems with a 
secondary M dwarf.  Table~\ref{tab:compmet} includes these values and Figure~\ref{fig:metcomp} shows the 
comparison. The ASPCAP metallicities are consistent with previous analysis; on average they are 0.07 dex more metal-rich with a scatter of 0.18 dex. Uncertainties
listed for these values in the literature do not always take into account systematic uncertainties in the abundance
analysis, which can be important when combining a heterogeneous set of metallicity derivations as is done
here. However, comparison of metallicities for well-studied stars in the literature, in particular for the
Gaia benchmark stars \citep{Jofre2014} show that the scatter there is typically $<0.1$ dex. Both of these effects are larger than the differences between the 25 high quality duplicate observations, which have difference of [M/H] = 0.007 and a scatter of 0.035 dex. Therefore, we conservatively
adopt 0.18 as the uncertainty in the ASPCAP metallicities and hope to be able to compare to a large set of
homogeneously derived high-resolution analyses in the future. For reference, we also
show in Figure~\ref{fig:metcomp} the comparison of our values to literature values that are derived from
low-resolution spectral indices calibrated to higher dispersion measurements.

\begin{table*}
\scriptsize
\centering
\caption{Comparison with Literature Metallicities} 
\label{tab:compmet} 
\begin{tabular}{llrrll}
\hline
2MASS ID  & Other ID & ASPCAP [M/H] & Literature [Fe/H] & Lit. method & Reference \\ 
\hline
2M01081597+5455148$^{\rm a}$   &     muCas     &  $-$0.82    &   $-$0.81  &  high-res & 1\\  
2M02043481+1249453 & & $-$0.54 & $-$0.41 & low-res & 2\\
2M02410716+5423087 & & 0.035& 0.61 & low-res & 2\\
2M03150093+0103083                   &	    NLTT10349   & $-$0.98  &	$-$0.92 &   binary   &	3\\	
2M03285302+3722579  	  &     LHS173      & $-$0.94    &$-$1.19 &   high-res & 4\\  
2M04342248+4302148 & & $-$0.09& 0.22 & low-res & 2\\
2M05011802+2237015  & & $-$0.91& 0.24 & low-res & 2\\
2M05312734$-$0340356    &	    GJ205      & 0.16  &        0.21   & high-res & 4\\ 
	    	      &	    	        & 0.16 & 	   0.35 &   low-res  &	 5\\
2M05454158+1107485  & & 0.04& 0.30 & low-res & 2\\
2M06181761+3200593  & & $-$0.07& 0.02 & low-res & 2\\
2M06312373+0036445    &      NLTT16628  & $-$0.50 &     $-$0.54 &   binary   &  3\\
2M06561894-0835461  & & $-$0.57& 0.10 & low-res & 2\\
2M08103429$-$1348514    &      GJ297.2B   & $-$0.33 &         0.03 &   binary   & 6\\
                      &     		&   &	$-$0.04 &   binary   &  7\\	
	    	      &	 		&   &	$-$0.04 &   low-res  &   5\\	
                      & &       & 0.01 & low-res & 2\\
2M08175130+3107455  & & $-$0.19& 0.27 & low-res & 2\\
2M08370799+1507475  & & $-$0.42& $-$0.11 & low-res & 2\\
2M08595755+0417552  & & $-$0.28& $-$0.10 & low-res & 2\\
2M09142298+5241125    &	    Gl338A	& $-$0.12&	$-$0.18	&   low-res  & 5\\
2M10112218+4927153    &     GJ380	& 0.02 &     $-$0.03 &   high-res & 4\\
  & &  &0.22 & low-res & 2\\
2M10335971+2922465  & & $-$0.01 & 0.03 & low-res & 2\\
2M10350859+3349499  & & $-$0.16 &$-$0.04 & low-res & 2\\
2M10361794+2844471  & & $-$0.31 &$-$0.18 & low-res & 2\\
2M10385685+2505402  & & $-$0.34 &$-$0.12 & low-res & 2\\
2M10453795+1833111  & & $-$0.08 & 0.14 & low-res & 2\\
2M10520440+1359509  & & $-$0.37 &$-$0.12 & low-res & 2\\
2M10550664+1532443  & & $-$0.23 &$-$0.05 & low-res & 2\\
2M10560279+4858238  & & $-$0.18 &0.02 & low-res & 2\\
2M11045698+1026411  & & $-$0.12 &0.00 & low-res & 2\\
2M11052903+4331357    &     GJ412A      & $-$0.60 &      $-$0.43 &   high-res & 4\\
	    	      &	    	&  &	$-$0.40	&   low-res  & 5\\
2M11091225-0436249  & & $-$0.26 &$-$0.03 & low-res & 2\\
2M11273856+0358359  & & 0.10 &0.47 & low-res & 2\\
2M11480063+3505146  & & $-$0.01 &0.21 & low-res & 2\\
2M11525880+3743060$^{\rm a}$     &     Gmb1830   & $-$1.31 &         $-$1.46  &  high-res & 1\\ 
2M11530522+1855480  & & $-$0.28 &$-$0.14 & low-res & 2\\
2M12192028+1323524  & & $-$0.04 &0.00 & low-res & 2\\
2M12210874+5642087  & & $-$0.47 &$-$0.40 & low-res & 2\\
2M12212146+5745089  & & $-$0.24 & 0.00 & low-res & 2\\
2M12241121+2653166  & & $-$0.22 &$-$0.13 & low-res & 2\\
2M12592744+5633464  & & $-$0.47 &$-$0.11 & low-res & 2\\
2M13095556+1438595  & & $-$0.53 &$-$0.21 & low-res & 2\\
2M13160127+1415504  & & $-$0.20 &$-$0.01 & low-res & 2\\
2M13315838+5443452  & & $-$0.22 & 0.09 & low-res & 2\\
2M13332256+3620352  & & 0.039  & 0.37 & low-res & 2\\
2M13514938+4157445  & & $-$0.14 & 0.39 & low-res & 2\\
2M13581901+0119475  & & $-$0.03 & 0.13 & low-res & 2\\
2M14045583+0157230    &     NLTT36190   & $-$0.27&      $-$0.03 &    binary   & 3\\ 
2M14050849+0312186  & & 0.23 & 0.52 & low-res & 2\\
2M14562809+1648342  & & $-$0.10 & 0.25 & low-res & 2\\
2M15202829+0011268    &     NLTT39942	& $-$0.25&      $-$0.38 &   binary   & 2 \\
2M16495034+4745402  & & $-$0.13 & 0.16 & low-res & 2\\
2M16535528+1138453  & & 0.12 & 0.62 & low-res & 2\\
2M17033253+1015052  & & $-$0.23 & $-$0.05 & low-res & 2\\
2M17190577+2253036  & & $-$0.02& 0.35 & low-res & 2\\
2M17592886+0318233  & & $-$0.14 & 0.03 & low-res & 2\\
2M18444674+4729496    &     KIC10318874	& 0.41 &	$-$0.12	&   low-res  & 8\\
2M18451027+0620158  & & $-$0.20 & 0.02 & low-res & 2\\
2M19081576+2635054  & & $-$0.84 & 0.39 & low-res & 2\\
2M19211069+4533525    &	    KIC9150827	& 0.01 &	$-$0.11	&   low-res  & 8\\	
2M19213157+4317347    &	    KIC7603200	& $-$0.31&	$-$0.18	&    low-res  & 8\\	
  & &  &$-$0.21 & low-res & 2\\
2M19283288+4225459    &	    KIC6949607	& 0.09&	$-$0.17	&   low-res  & 8\\	
2M19300081+4304593    &	    KIC7447200	& 0.22&	$-$0.12	&    low-res  & 8\\	
2M19312949+4103513    &	    KIC5794240	& 0.11 &	   0.20	&    low-res  & 8\\	
2M19343286+4249298    &	    KIC7287995	& $-$0.08&	$-$0.20	&   low-res  & 8\\	
2M19513233+0453486  & & $-$0.06 & 0.31 & low-res & 2\\
2M21105737+4657578  & & $-$0.18 & 0.22 & low-res & 2\\

\hline
\end{tabular}
\begin{list}{}{}
\item $^{\rm a}$ Outside temperature range, but included as a GAIA-ESO calibration star
\item References---1) \citet{Jofre2014}; 2) \citet{Terrien2015}; 3) \citet{mann13}; 4) \citet{Woolf2005}; 5) \citet{Rojas-Ayala2012}; 6) \citet{Neves2012};  7) \citet{Fuhrmann2008}; 8) \citet{Muirhead2012}
\end{list}
\end{table*}

\section{Photometry}
\label{sec:phot}

To test the relationships between metallicity, $T_{\rm eff}$, and colour, we combine the APOGEE parameters with photometry 
from SDSS, 2MASS, and WISE. Photometry from each of these surveys is included in Table~\ref{tab:all}.

\subsection{SDSS}
\label{sec:sdss}

In general, APOGEE targets are saturated in SDSS photometry, because the H$<$12.2 mag
limit for most APOGEE observations means that the $ugriz$ magnitudes are too bright
for the $\sim$14.5 magnitude bright limit for the SDSS photometric survey. However,
for the reddest stars, particularly in the deeper ($H <13.8$) APOGEE fields, there
are stars with good measurements in both surveys, including stars deliberately targeted
by APOGEE as overlap targets with the SEGUE optical spectroscopic survey \footnote{We note that while there are stars with data from both surveys, the SEGUE pipeline only produces reliable parameters for warmer stars \citep[$T_{\rm eff} > 4500$~K][]{Lee2008} so SEGUE parameters are not useful calibrators for this APOGEE sample.}.

SDSS photometry was obtained from the Data Release 10 \citep[DR10;][]{dr10} database via a
coordinate cross-match using the online object cross-ID.\footnote{\url{http://skyserver.sdss3.org/dr10/en/tools/crossid/crossid.aspx}} 
Each APOGEE KM star was matched to the closest SDSS photometric point source within 5\arcsec. 
The APOGEE fields do not entirely overlap with the SDSS photometry footprint, so only 
2977 of the 4246 total APOGEE stars had matches in the DR10 photometric database. 
To select stars with good photometry, we performed cuts using both the photometric flags and quoted photometric uncertainties. 
Our SDSS flag cuts are based on the SDSS photometric flag recommendations\footnote{\url{http://www.sdss.org/dr12/algorithms/photo_flags_recommend/}}, implemented to exclude only the 
band where the flags indicate poor photometry. The flags we used are listed in 
Table~\ref{tab:sdssflags} with the number of objects with the flags triggered in each 
band. The majority of the bad photometry was saturated; these saturated stars usually 
also triggered flags for poor interpolation (psf\_flux\_interp and interp\_center). The remaining bad photometry was due to objects located on the edges of images and blends with nearby objects. 

\begin{table} 
\centering
\caption{SDSS Photometric Flags} 
\label{tab:sdssflags} 
\begin{tabular}{llllll}
\hline
Flag  & $u$ & $g$ & $r$ &  $i$ & $z$ \\ 
\hline
edge  &  43  &  58  &  58  &  60  &  42 \\ 
satur  &  7  &  159  &  772  &  1363  &  163 \\ 
nodeblend  &  133  &  133  &  133  &  133  &  133 \\ 
peakcenter  &  12  &  17  &  64  &  128  &  14 \\ 
notchecked  &  31  &  64  &  119  &  115  &  37 \\ 
dblend\_nopeak  &  12  &  7  &  14  &  22  &  6 \\ 
psf\_flux\_interp  &  68  &  145  &  759  &  1352  &  177 \\ 
bad\_counts\_error  &  1  &  1  &  8  &  11  &  2 \\ 
interp\_center  &  99  &  160  &  771  &  1370  &  193 \\ 
\hline
total rejected & 301 & 404 & 985 & 1573 & 394 \\ 
total good & 2676 & 2573 & 1992 & 1404 & 2583 \\
\hline
\end{tabular}
\end{table}

The flag cuts discarded over half of the $i$ band photometry, but included a larger fraction of detections in the $u$, $g$, $r$, and $z$ bands. After discarding the flagged photometry, we selected uncertainty cuts for each band based on the error distribution; we first fit each distribution by a gaussian, then selected the mean of the gaussian plus two times the standard deviation as the highest uncertainty included in the data. The resulting uncertainty cuts and the number of detections passing them in each band are given in Table~\ref{tab:phot_unc}. Extinction corrections are discussed in Section~\ref{sec:ext}

\begin{table} 
\centering
\caption{SDSS-2MASS-WISE numbers and uncertainty limits} 
\label{tab:phot_unc} 
\begin{tabular}{llllll}
\hline
band & \# & \# & $\sigma$ & \# & \# \\
 & initial & good & uncertainty  & passing & with good \\
 & match & phot &  limit & error cuts & extinction  \\
\hline
$u$   & 2977 & 2676 & 0.035 & 2077 & 2038 \\
$g$   & 2977 & 2573 & 0.026 & 2264 & 2159 \\
$r$   & 2977 & 1992 & 0.021 & 1732 & 1634 \\
$i$   & 2977 & 1404 & 0.022 & 1232 & 1155 \\
$z$   & 2977 & 2583 & 0.023 & 2284 & 2179 \\
$J$   & 4246 & 4155 & 0.024 & 3595 & 3282 \\
$H$   & 4246 & 4123 & 0.030 & 3523 & 3216 \\
$K_S$ & 4246 & 4148 & 0.025 & 3579 & 3280 \\
$W1$  & 4206 & 3824 & 0.024 & 2985 & 2795 \\
$W2$  & 4206 & 3744 & 0.025 & 3372 & 3125 \\
$W3$  & 4206 & 3795 & 0.10  & 1159 & 1116 \\
$W4$  & 4206 & 441 & 0.20  & 72 & 70 \\
\hline
\end{tabular}
\end{table}

\subsection{2MASS}
\label{sec:2mass}
While 2MASS photometry was used to selected APOGEE targets and is included in the database, we obtained photometry from the 2MASS All-Sky Point Source catalog to ensure consistent flag and uncertainty cuts. All 4246 stars had matches in the point source catalog within 5\arcsec. Flag cuts were performed on each band individually (instead of cutting all three bands if one was poor) to include the largest possible sample of good photometry. We required each band have reliable photometry (ph\_qual=ABCD), contain no saturated pixels (rd\_flg=2), be either unblended or be properly deblended (bl\_flg$>0$), and be uncontaminated by artifacts (cc\_flg=0). Our uncertainty cuts, selected using the same method as those for SDSS, are given in Table~\ref{tab:phot_unc}. 

\subsection{WISE}
\label{sec:wise}
We obtained WISE photometry from the ALLWISE catalog via a coordinate cross-match within 5\arcsec, obtaining matches for 4206 of the 4246 total objects. Flag cuts were again performed on each band;  we required each band to be marked as reliable photometry (ph\_qual=ABC), uncontaminated (cc\_flags = 0), not part of an extended source (ext\_flg$<$2), relatively uncontaminated by the moon (moon\_lev $<$5) and less than 20\% saturated. The uncertainty cuts are given in Table~\ref{tab:phot_unc}. The majority of KM dwarfs with WISE matches have reliable $W1$ and $W2$ magnitudes but not $W3$ and $W4$ magnitudes, due to the much brighter limits on the further infrared bands (the 95\% completeness levels are $W1<17.1$, $W2<15.7$, $ W3<11.5$, and $W4<7.7$\footnote{\url{http://wise2.ipac.caltech.edu/docs/release/allwise/expsup/sec2_1.html}}).

\subsection{Extinction}
\label{sec:ext}
While the APOGEE KM sample consists of relatively nearby stars ($d<600$~pc), extinction due to Galactic dust can alter the colours of objects more distant than $d\sim50$~pc \citep{Leroy1993}, especially those that fall outside the local bubble \citep[$d\sim100$~pc; e.g.,][]{Lallement2003,Jones2011}. Extinction maps designed for extra-galactic studies \citep[e.g.,][]{Schlegel1998} overestimate the extinction for these nearby dwarfs, but three-dimensional maps require accurate distances that are not available for these low mass stars at sub-solar metallicity. To estimate distances, we first calculated stellar radii from the \citet{Mann2015} coefficients based on $T_{\rm eff}$ and [M/H], then calculated $K_S$ magnitudes using the \citet{Mann2015}
 radius-metallicity-magnitude relation. We eliminated 22 M dwarfs in our sample with [M/H] $< -1$ because the relationships were not calibrated for these low-metallicity stars. We then calculated distances based on the difference between estimated and observed $K_S$ magnitudes. 

We obtained $E(B-V)$ extinction values from a three dimensional dust map based on a combination of Pan-STARRS and 2MASS data \citep{Schlafly2014,Green2015}. The map is presented in integer and half-integer values of distance modulus, so for each star we queried the online database to obtain a minimum and maximum extinction from those discrete values. The incomplete overlap of APOGEE and Pan-STARRS excluded 36 APOGEE KM dwarfs. The \citet{Green2015} extinction values were calculated based on the color difference between foreground and background stars, so in for the nearest stars (closer than $d\sim100$~pc) they include extrapolated values based on a Galactic model. For 221 stars with estimated distances less than $d<50$~pc, we assume $E(B-V)=0$, and for 3967 stars further than $d>50$~pc, we adopt the \citet{Green2015} values for minimum and maximum $E(B-V)$. 

The resulting extinction values have a median $E(B-V)=0.01$ and a median difference between minimum and maximum of $\Delta E(B-V)=0.002$. We exclude 290 stars with $E(B-V)>0.1$ and 64 stars with $\Delta E(B-V)>0.02$ because those stars are located in or near dust clouds and our distances are not precise enough to accurately estimate their extinctions. Stars without accurate extinctions are excluded from the APOGEE KM sample, resulting in a final sample of 3834. Due to the relatively small reddening values for the APOGEE KM sample, we did not adjust our estimated distances by using the apparent magnitude
corrected for the reddening and iterating until convergence. The final number of stars with photometry in each band is given in Table~\ref{tab:phot_unc}.

We calculated the $A_{\lambda}$ values for the SDSS $ugriz$ and APASS $V$ using the $R=3.1$ extinction law from \citet{Schlafly2011}. To extinction correct the 2MASS $JHK_S$ and WISE $W1W2$, we converted $A_r$ to $A_{\lambda}$ values using the relationships from \citet{Davenport2014}. We do not correct $W3$ and $W4$ photometry both because the corrections are well below the uncertainties on the magnitudes in both bands and because the number of KM dwarfs with reliable photometry in those bands is small. The values presented in Table~\ref{tab:phot_unc} and used throughout the paper have been corrected for extinction.

\section{Stellar Model Isochrones}
\label{sec:starmods}
The accurate $T_{\rm eff}$ and [M/H] values measured from APOGEE data provide a unique opportunity to test the relationships between colour, $T_{\rm eff}$, and metallicity as compared to stellar isochones. We examine these relationships in comparison with three model grids: Dartmouth \citep{dotter08}, PARSEC \citep{Bressan2012}, and BT-Settl \citep{allard03,Allard2011}. In each set of models, we selected a single 2~Gyr isochrone for comparison. This is a good match for the mean age of nearby field stars, but the photometry of late-K and early-M (3500~K~$\le T_{\rm eff} \le 4200$~K) dwarf stars is not sensitive to the age choice between 0.1 and 10~Gyr so we expect the single isochrone to be a good match for the range of ages. 

The KM dwarfs in the solar neighborhood do have a range of [$\alpha$/Fe] enhancement, including both [$\alpha$/Fe]-rich ($> 0.2$dex) and [$\alpha$/Fe]-poor ($< 0.2$ dex) stars \citep[e.g.,][]{Bensby2003,Adibekyan2012} in the range $-1 < $[Fe/H]$< -0.3$. The [$\alpha$/M] values reported by the ASPCAP pipeline for the KM stars in our sample show a similar bimodality, although there are no literature values for our sample of stars to test the accuracy of individual stellar measurements for this abundance ratio. Therefore, we have not added [$\alpha$/M] as an additional parameter at the present time. Instead, in our comparisons with the model grids below, we show both $\alpha$-poor and $\alpha$-rich versions where possible. For each model grid, we selected isochrones based on the \citet{caffau11} solar abundances. 

The BT-Settl model isochrones are based on the stellar evolution codes of \citet{baraffe97,Baraffe1998,Chabrier1997} with an updated version of the PHOENIX stellar atmosphere code \citep{Hauschildt1999} that is optimized for low mass stars and dusty brown dwarfs \citep{allard03,Allard2011}. We retrieved photometry in the SDSS, 2MASS, and WISE bands for isochrones that span [M/H] from $-1.0$ to $0.0$ with a spacing of 0.5 dex.\footnote{\url{https://phoenix.ens-lyon.fr/Grids/BT-Settl/CIFIST2011bc/}} The only available [$\alpha$/Fe] is scaled in an approximation of the thin disk, assuming [$\alpha$/F] = 0.0 for [M/H] = 0, [$\alpha$/Fe] = 0.2 for [M/H] = -0.5, and [$\alpha$/Fe] = 0.4 for [M/H] $=-1.0$.

The PARSEC models are the most up-to-date result from the Padova and Trieste stellar evolution codes \citep{Bressan2012,Chen2014}. For comparison to the APOGEE KM photometry, we chose the 1.2S models\footnote{Available from the CMD 2.7 input form \url{http://stev.oapd.inaf.it/cmd}} converted from luminosities to SDSS, 2MASS, and WISE photometry using bolometric corrections derived based on the BT-Settl atmospheres \citep{Chen2014}. We included no interstellar reddening or circumstellar dust. The PARSEC metallicities are given in terms of $Z$ and are fixed at a scaled solar abundance, so we could not investigate $\alpha$-enhanced versions of these isochrones. We converted $Z$ to [M/H]=[Fe/H] via the relation [M/H]$ = \log(Z/Z_{\sun})$, using the \citet{caffau11} value of $Z_{\sun} = 0.0152$. To compare to the APOGEE KM sample, we downloaded two tracks with metallicities of [M/H]=[Fe/H] =0.0 and  [M/H]=[Fe/H] =$-$0.7. 

\citet{dotter08} presented the Dartmouth Stellar Evolution Database, which contains models from the Dartmouth Stellar Evolution Program
and additional software tools. The Dartmouth Isochrones are translated from the evolutionary models using the PHOENIX stellar atmosphere code \citep{Hauschildt1999}. We used the Dartmouth Isochrone and LF Generator\footnote{\url{http://stellar.dartmouth.edu/models/isolf_new.html}} to obtain isochrones in SDSS, 2MASS, and WISE photometry. We adopted the default helium abundance of $Y = 0.245 + 1.5 \cdot Z$ and used cubic interpolation to construct our model grid. We retrieved models with abundances to match both the BT-Settl and PARSEC abundances, including [Fe/H]=0 and [$\alpha$/H] = 0.0, [Fe/H]=$-$0.7 and [$\alpha$/H] = 0.0, [Fe/H]=$-$0.7 and [$\alpha$/H] = 0.2, and [Fe/H]=$-$1.4 and [$\alpha$/H] = 0.4.

\section{Colours as Indicators of Temperature and Metallicity}
\label{sec:compare}
K and M dwarfs with $T_{\rm eff}$ from 3550 to 4200~K are some of the most numerous stars, but the link between their metallicities and broad-band colours is poorly understood. Our APOGEE KM sample presents a unique opportunity to examine SDSS-2MASS-WISE colours sensitive to $T_{\rm eff}$ and [M/H]. Table~\ref{tab:all} includes the collected photometry and ASPCAP parameters for the stars used in this analysis.

\subsection{Relationships between colour and $T_{\rm eff}$}
\label{sec:modcomp}
\begin{figure*}
\includegraphics[width=0.85\linewidth]{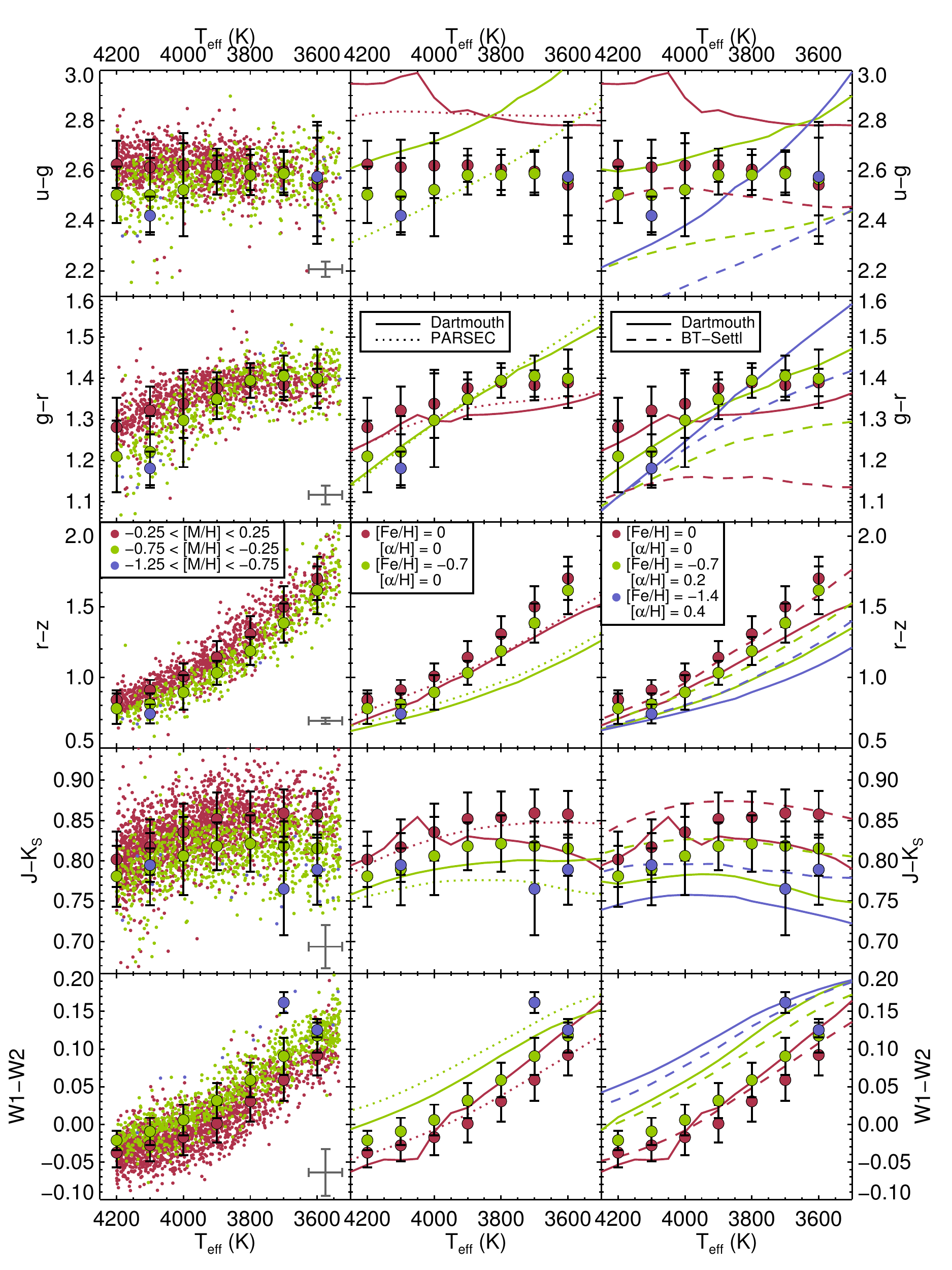}
\caption{The $u-g$, $g-r$, $r-z$, $J-K_S$, and $W1-W2$ colours of APOGEE KM dwarfs as a function of $T_{\rm eff}$. The left row of panels show data for individual stars, while the middle and left panels show only the mean and standard deviation. The middle panels show APOGEE KM colours compared to Dartmouth and PARSEC models with scaled solar abundances, while the right panels show APOGEE KM colours compared to Dartmouth and BT-Settl colours with $\alpha$ enhanced abundance patterns.}
\label{fig:coltemp_mod}
\end{figure*}

In Figure~\ref{fig:coltemp_mod}, we show five representative colours as a function of $T_{\rm eff}$\footnote{Additional colour/$T_{\rm eff}$ relations can be explored using \url{https://filtergraph.com/apogee_km_dwarf_colors}} compared to the colours of the model grids described in Section~\ref{sec:starmods}.  

For KM dwarfs with $T_{\rm eff} > 3900$~K, we find that metal-poor stars have bluer $u-g$ and $g-r$ colours than more metal-rich stars. This is consistent with the pattern for G and K stars, which have blue $u-g$ and $g-r$ colours due to decreased opacity from metal lines in the bluest bands \citep{Roman1954}. For the cooler stars in our sample ($3500 < T_{\rm eff} < 3900$~K), we find instead that $u-g$ and $g-r$ are constant as a function of both $T_{\rm eff}$ and metallicity. This is not completely consistent with previous results for early-M dwarfs, which indicate that metal-poor M dwarfs have redder $g-r$ colours than their solar metallicity counterparts \citep{West2004,Lepine2008,Bochanski2013}, tracing a similar effect in to that observed in $B-V$ \citep{Gizis1997} and $B-R$ \citep{Hartwick1977}. The APOGEE KM sample indicates that the transition from blue metal-poor stars to red metal-poor stars must happen at $T_{\rm eff}<3500$~K ($\sim$M2), while each model grid indicates the transition occurs for stars hotter than $T_{\rm eff}=4000$~K (the transition between K and M dwarfs). The models are generally a poor match for the data in both $u-g$ and $g-r$, and applying the 130~K offset derived in Section~\ref{sec:teff} would not improve the agreement. 

The $r-z$ colour shows a strong dependence on $T_{\rm eff}$, as has been demonstrated by previous SDSS work \citep{Bochanski2010,Dhital2010}. This $T_{\rm eff}$ dependence mimics behavior previously observed in $r-K_S$ and $V-K$. In addition to the correlation with temperature, $r-z$ shows an offset in metallicity, also recently examined by \citet{Bochanski2013} using the statistical parallax technique. All three model grids reproduce this offset between metal-poor and solar metallicity stars, but there is a shift between the models and data in colour/$T_{\rm eff}$ space. If the 130~K offset derived in Section~\ref{sec:teff} was applied, the overlap between models and data would be significantly better for all isochrones.

For these KM dwarfs, the $J-K_S$ (in addition to $J-H$ and $H-K_S$, not shown) colour varies weakly with spectral type/$T_{\rm eff}$ \citep[e.g.,][]{Covey2007,Davenport2014}. The main variable driving variation in these infrared bands is [M/H], a dependence that has been examined for M dwarfs \citep{Leggett1992,Johnson2012,Newton2014}. The BT-Settl model grid matches the $J-K_S$ colour relatively well, while the Dartmouth and PARSEC grids only reproduce the magnitude of the shift between metallicity bins and not the actual colours. Applying the $130$~K offset discussed in Section~\ref{sec:teff} would not result in better agreement between data and models. 

The $W1-W2$ colour is correlated with both metallicity and temperature, with metal-poor stars having redder colours than their solar metallicity counterparts. While the offset is not large compared to typical WISE photometric uncertainties, because metal-poor stars are red instead of blue, $W1-W2$ can be a useful way of disentangling $T_{\rm eff}$ and [M/H] using photometry alone (see Section~\ref{sec:relations}). The models generally reproduce the redder color of metal-poor stars, but do not produce the observed relationship between colour and $T_{\rm eff}$. In contrast with $r-z$, the disagreement between the models and the data would become worse if the $130$~K offset derived in Section~\ref{sec:teff} was applied. 

\subsection{Metallicity in colour-colour space}
\label{sec:colcol}
\begin{figure*}
\includegraphics[width=\linewidth]{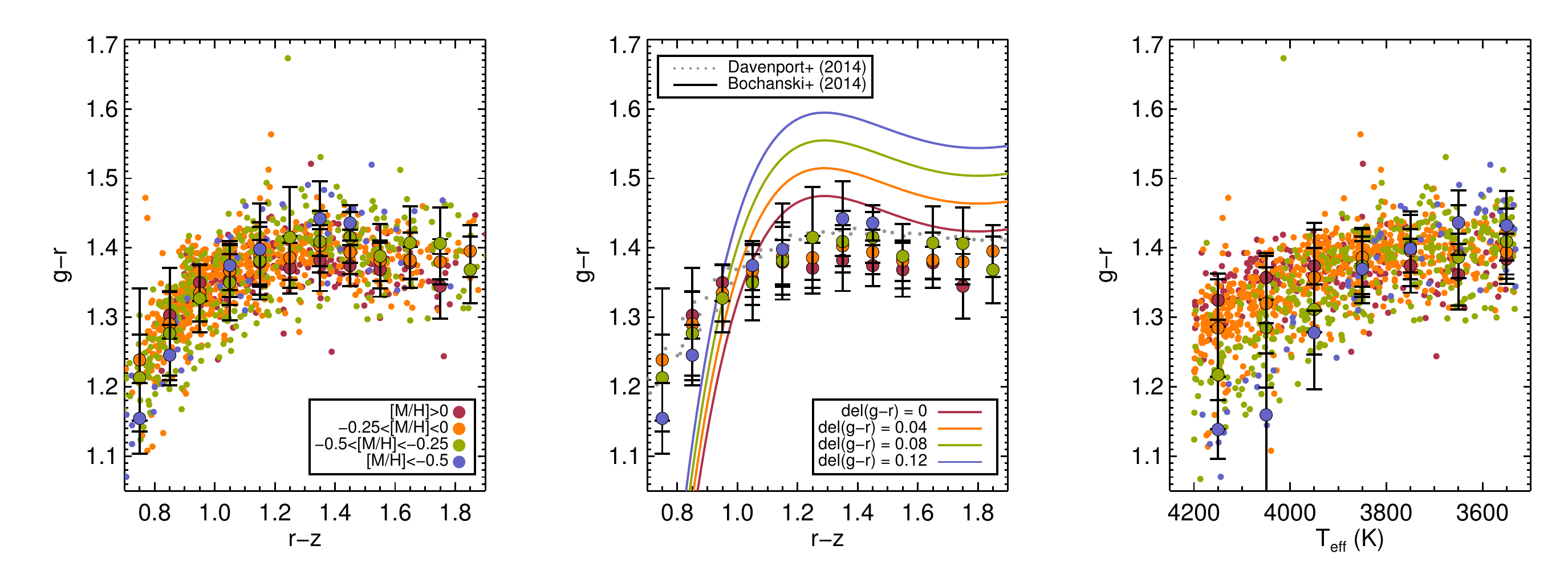}
\includegraphics[width=\linewidth]{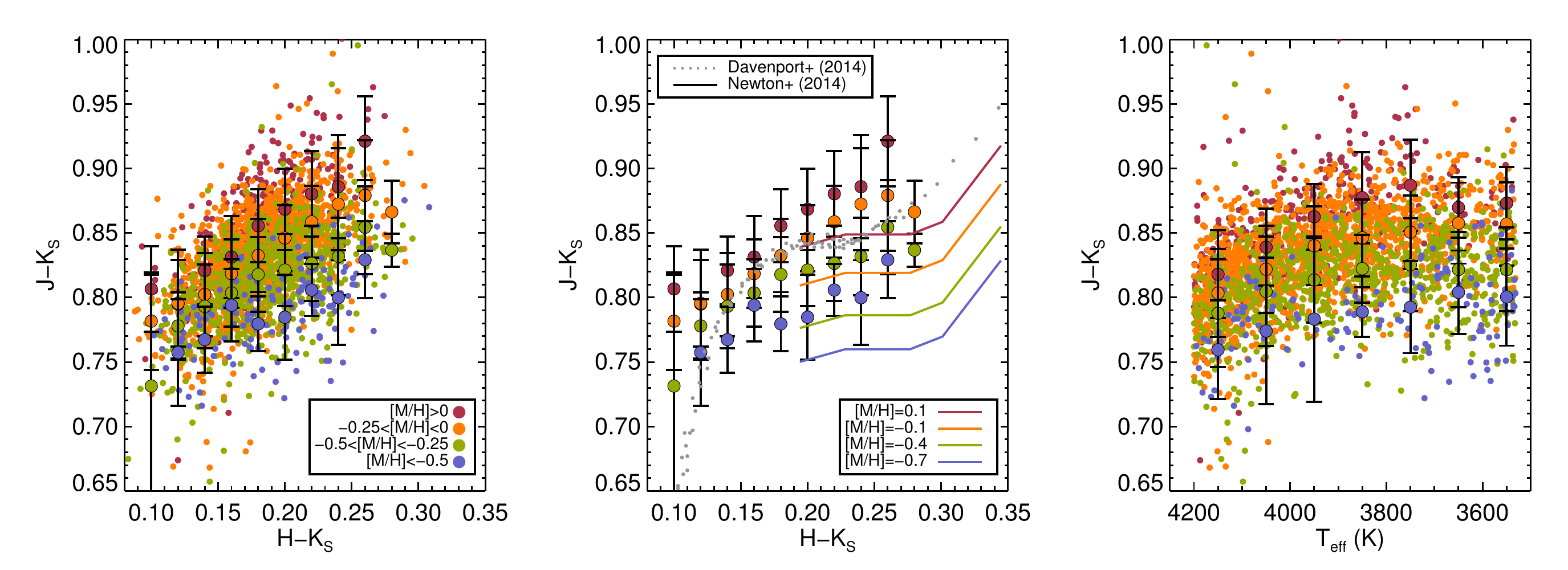}
\caption{\textbf{Top Row:}  $g-r$ colour as a function of $r-z$ colour (left and center panels) and $T_{\rm eff}$ (right panel) for the APOGEE KM sample, with colour indicating metallicity bin. The median and standard deviation $g-r$ colours in each metallicity bin are shown in bins of 0.1 in colour and 100~K $T_{\rm eff}$. In the center panel, we also show the \citet{Davenport2014} colour-colour locus and the \citet{Bochanski2013} polynomial fit to $g-r$ as a function of $r-z$ at solar metallicity, in addition to lines for $\delta(g-r)=0.04$, 0.08, and 0.12. If a linear relationship between $\delta(g-r)$ and [M/H] can be assumed, the line colours should match to the metallicity bins; based on the lack of agreement between the $\delta(g-r)$ relationship and the APOGEE KM data, a linear relationship is unlikely. \textbf{Bottom Row:} $J-K_S$ colour as a function of $H-K_S$ colour (left and center panels) and $T_{\rm eff}$ (right panel) for the APOGEE KM sample, with colour indicating metallicity bin.  The median and standard deviation $g-r$ colours in each metallicity bin are shown in bins of 0.02 in colour and 100~K $T_{\rm eff}$. In the center panel, we also show the \citet{Davenport2014} colour-colour locus and the \citet{Newton2014} metallicity relation are also shown. The reddest and bluest bins for the APOGEE KM sample are likely incomplete.}
\label{fig:colcol}
\end{figure*}

Only two of the photometric metallicity relations derived for KM dwarfs overlap with SDSS-2MASS-WISE photometry used to examine the APOGEE KM sample, a $g-r$/$r-z$  relation from \citet{Bochanski2013} and a $J-K_S$/$H-K_S$ relation from \citet{Newton2014}. Figure~\ref{fig:colcol} shows the KM sample in these two colour-colour spaces compared to both these relations. 

\citet{Bochanski2013} use $\delta(g-r)$ as a proxy for metallicity for M dwarfs, following on previous work \citep{West2004,West2011} showing that M subdwarfs (classified based on low-resolution optical spectra) have red $g-r$ colours. \citet{Bochanski2013} fit a polynomial to solar-metallicity M dwarfs in $g-r$ as a function of $r-z$, then divide M dwarfs into increasingly metal-poor bins based on $\delta(g-r)$, which quantifies how much bluer each M dwarf is than the solar-metallicity stars. While the \citet{Bochanski2013} $\delta(g-r)$ is correlated to metallicity \citep[as shown through its correlation with $\zeta$, the low-resolution spectroscopic metallicity parameter;][]{Dhital2012,Lepine2007} it is not explicitly calibrated to metallicity. To compare the APOGEE KM data with the \citet{Bochanski2013} $\delta(g-r)$, we used a rough equivalence between  $\delta(g-r)$ and $\zeta$, then the $\zeta$ metallicity relation of \citet{Woolf2009}. 

The agreement is poor between the \citet{Bochanski2013} relation and the APOGEE KM data. The \citet{Bochanski2013} solar-metallicity line does not overlap with the bulk of the APOGEE KM data. The APOGEE KM colours match well with the fiducial colour locus for field stars \citep{Davenport2014}, so the \citet{Bochanski2013} solar metallicity fit was probably based on incomplete data for $r-z<1$. The $\delta(g-r)$ values also show a strong relationship between colour and metallicity, while the data show $g-r$ is only weakly dependent on metallicity. The $\delta(g-r)$ metallicity indicator may not be useful for stars near solar metallicity, and should instead be restricted to the more metal-poor stars ([Fe/H]$ < -1$; subdwarfs and extreme subdwarfs) that were used to derive the indicator. 

\citet{Newton2014} use low-resolution infrared lines as metallicity indicators to calibrate a relation based on 2MASS $J-K_S$ and $H-K_S$ colours, which is compared to the APOGEE KM data in the bottom row of Figure~\ref{fig:colcol}. The \citet{Newton2014} calibration relies on the \citet{Bessell1988} colour-colour locus \citep[translated to 2MASS bands using ][]{Carpenter2001}, then fits an offset in $J-K_S$ to metal-poor M dwarfs with blue 2MASS $J-K_S$ colours. Again, the agreement between the solar-metallicity line and our colours is poor, but the lack of agreement between the APOGEE KM dwarfs and the \citet{Davenport2014} locus indicates that the reddest and bluest bins in $H-K_S$ are likely to be biased and/or incomplete for the APOGEE KM sample. This is reasonable, as $H-K_S$ is sensitive to both metallicity and $T_{\rm eff}$ and the sample is selected based on $T_{\rm eff}$. Despite the poor agreement between the solar metallicity locus in colour-colour space, both samples show the same change in $J-K_S$ colour due to metallicity (i.e. [M/H] = 0 stars are consistently $\delta(J-K_S)\sim0.7$ redder than [M/H] = 0.5). 

\subsection{Empirical $T_{\rm eff}$ and [M/H] relations}
\label{sec:relations}
While every colour is sensitive to both $T_{\rm eff}$ and [M/H], $r-z$ and $W1-W2$ are particularly good tracers of both physical properties. Both $r-z$ and $W1-W2$ become redder as a function of $T_{\rm eff}$ and show a strong shift in colour between solar metallicity and metal-poor stars. Classic SDSS [M/H] indicators ($u-g$ and $g-r$) are degenerate for $3600 < T_{\rm eff} < 3800$~K and so are not ideal for these objects, and it is difficult to separate the $T_{\rm eff}$ and [M/H] dependence of 2MASS colours ($J-H$, $J-K_S$) because their relationship with $T_{\rm eff}$ is weak and non-linear. Because metal-poor stars are blue in $r-z$ and red in $W1-W2$, the [M/H]/$T_{\rm eff}$ space is non-degenerate in this color combination. We provide fits for [M/H] and $T_{\rm eff}$ as a function of both $r-z$ and $W1-W2$.

\begin{figure*}
\includegraphics[width=\linewidth]{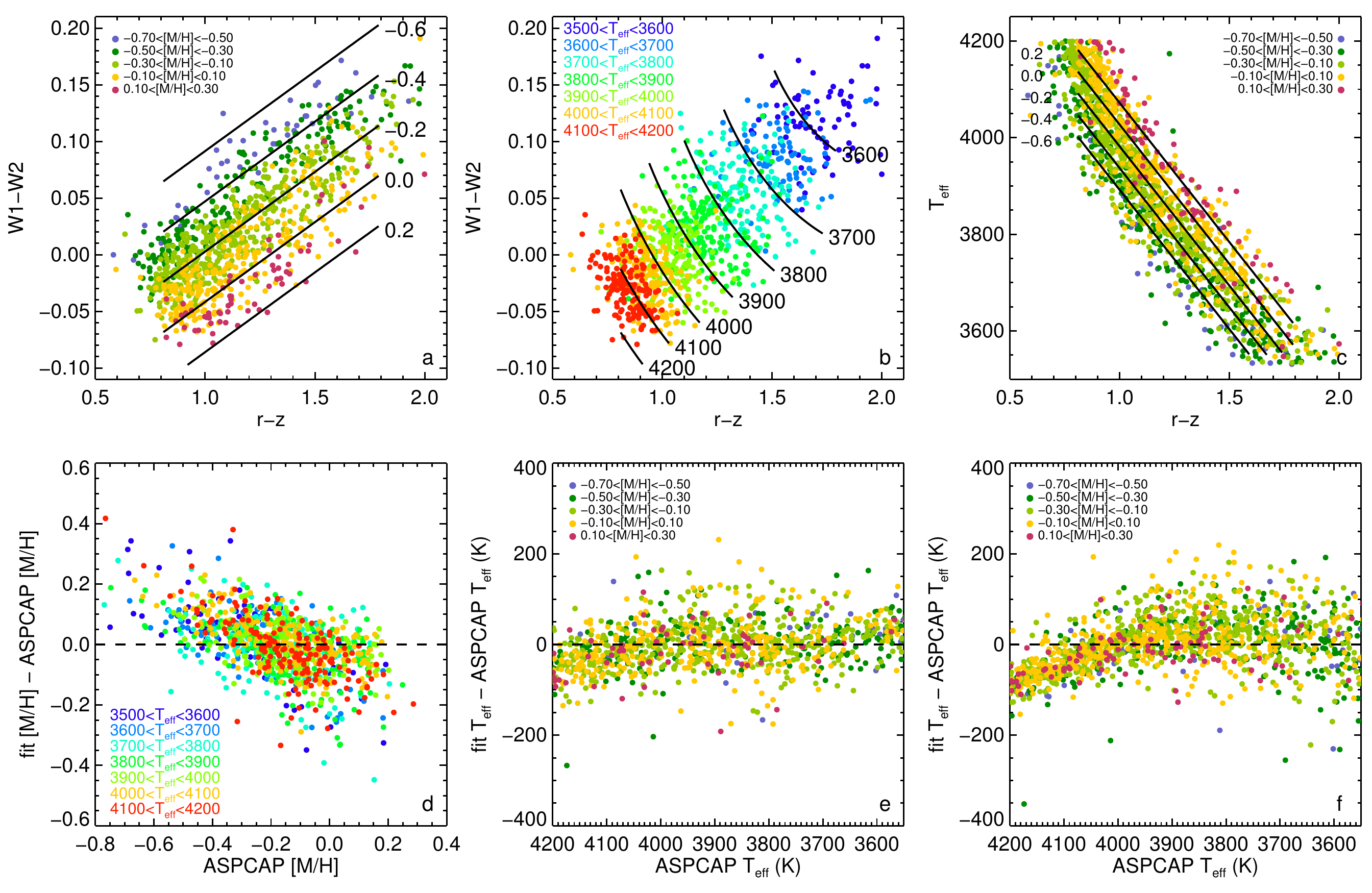}
\caption{Relationships between colour, [M/H], and $T_{\rm eff}$ and fit uncertainties for the corresponding fits. Panel a (top left) shows $W1-W2$ as a function of $r-z$ colour with points coloured according to their [M/H].  Lines of constant [M/H] calculated using the multi-dimensional fit are shown, and the difference between fit [M/H] and ASPCAP [M/H] as a function of ASPCAP [M/H] is shown in panel d (bottom left) with points colour coded according to $T_{\rm eff}$. Panel b (top middle) shows $W1-W2$ as a function of $r-z$ colour with points coloured according to their $T_{\rm eff}$. Lines of constant $T_{\rm eff}$ calculated using the multi-dimensional fit are shown, and the difference between fit $T_{\rm eff}$ and ASPCAP $T_{\rm eff}$ as a function of ASPCAP $T_{\rm eff}$ is given in panel e (bottom middle), colour coded according to [M/H]. Panel c (top left) shows $T_{\rm eff}$ as a function of $r-z$ colour with stars colour-coded based on their [M/H]. The fit to $T_{\rm eff}$ as a function of $r-z$ and [M/H] is shown for labeled values of [M/H]. Panel f (bottom right) shows difference between the $T_{\rm eff}$ calculated from the fit and the ASPCAP $T_{\rm eff}$ value as a function of ASPCAP $T_{\rm eff}$ with points coloured according to [M/H]. }
\label{fig:coeff}
\end{figure*}

The relationship between [M/H] and these two colors is shown in panel a of Figure~\ref{fig:coeff}. The coefficients for the fit to [M/H] as a function of $r-z$ and $W1-W2$ are given in Table~\ref{tab:coeff}, and the fit is shown with lines of constant [M/H] in panel a of Figure~\ref{fig:coeff}. Panel d of Figure~\ref{fig:coeff} shows the fit [M/H] compared to the ASPCAP values. The fit shows a systematic trend with [M/H] but no trend with $T_{\rm eff}$; the fit is not improved by increasing the polynomial degree. We adopt the scatter in difference between the fit and measured values (0.1 dex) as the uncertainty as it is significantly larger than the formal errors on the fit.

\begin{table*} 
\centering
\caption{Coefficients for the relations between colour, $T_{\rm eff}$, and [M/H]} 
\label{tab:coeff} 
\begin{tabular}{lllllll}
\hline
$Y^a$ & $a_0$ &  $a_1$ &  $a_2$ & $a_3$ &  $a_4$ & $\sigma$ \\
\hline
[M/H] & $-$0.822 & 0.634 & \nodata & $-$4.508 & \nodata  & 0.102  \\
$T_{\rm eff}^c$ & 4707.2 & $-$958.0 & 226.6 & $-$1554.2 & 2849.7 & 53.3  \\ 
\hline
\hline
$Y^b$ & $b_0$ &  $b_1$ &  $b_2$ & & & $\sigma$ \\
\hline
$T_{\rm eff}^c$ & 4603.4 & $-$576.5 & 225.0 & & & 61.5  \\ 
\hline
\multicolumn{7}{l}{All relations are only valid $0.8 < r-z < 1.8$} \\
\multicolumn{7}{l}{$^a$ Form $Y = a_0 + a_1(r-z) + a_2(r-z)^2 + a_3(W1-W2) + a_4(W1-W2)^2$} \\
\multicolumn{7}{l}{$^b$ Form $Y = b_0 + b_1(r-z) + b_2[M/H]$} \\
\multicolumn{7}{l}{$^c$ ASPCAP $T_{\rm eff}$ values were not corrected to match interferometric values}
\end{tabular}\\
\end{table*}

The $T_{\rm eff}$ values as a function of $r-z$ and $W1-W2$ are shown in panel b of Figure~\ref{fig:coeff}. This relationship was best fit by a second degree polynomial in both colours; the coefficients are given in Table~\ref{tab:coeff} and the fit is shown with lines of constant $T_{\rm eff}$ in panel b of Figure~\ref{fig:coeff}. The difference between fit $T_{\rm eff}$ and the ASPCAP values is shown in panel e of Figure~\ref{fig:coeff}. The scatter in the difference is $\sigma=53$~K, and there are no systematic effects in $T_{\rm eff}$ or [M/H]. While in Section~\ref{sec:teff} we found that ASPCAP $T_{\rm eff}$ is 130~K hotter than the interferometic-based $T_{\rm eff}$ values of \citet{Boyajian2013} and \citet{Mann2015}, we did not apply this offset to our $T_{\rm eff}$ values before performing the fit. 

We also provide a fit for $T_{\rm eff}$ as a linear function of both $r-z$ and [M/H] for objects with [M/H] measured from other sources.\footnote{A similar fit for $W1-W2$ did not provide accurate $T_{\rm eff}$ values.} The data and associated fit are shown in panel c of Figure~\ref{fig:coeff}, and the coefficients are given in Table~\ref{tab:coeff}. The difference between ASPCAP $T_{\rm eff}$ and fit $T_{\rm eff}$ is shown in panel d of Figure~\ref{fig:coeff}. The linear fit is poor at the high $T_{\rm eff}$ end ($T_{\rm eff}>4100$~K), but higher order polynomials did not provide a better fit in that temperature regime. Due to the systematics and the slightly higher scatter ($\sigma=62$~K), this relation should be used only if $W1-W2$ photometry is unavailable. Despite the low dispersion on the fits, the precision of these relations is limited uncertainties on the data ($\sigma=100$~K and $\sigma=0.18$~dex).

\section{Conclusions}
\label{sec:conclusions}
Precise and accurate determinations of [M/H] and $T_{\rm eff}$ of late-K and early-M dwarfs based on photometric indicators, especially when combined with upcoming Gaia parallaxes, will revolutionize our understanding of early Galactic evolution. As APOGEE continues to obtain spectra and update the associated model grids, we expect larger numbers of low mass stars with more accurate parameters. The combination of those values with colours will be very powerful.

We determined that the current ASPCAP catalog \citep{garcia2015} includes parameters for late-K and early-M dwarfs with $T_{\rm eff}$ accurate to 100~K (with a 130~K offset) and [M/H] accurate to 0.18 dex. Using those values for the APOGEE KM sample, we examined the relationship between colour, $T_{\rm eff}$, and [M/H] across SDSS, 2MASS, and WISE bands. We find that nearly every colour shows some sensitivity to $T_{\rm eff}$ and [M/H], though we note that $g-r$ is not very sensitive to metallicity for the cool end of our sample ($T_{\rm eff} < 3900$~K; corresponding to M0--M2 dwarfs). We confirm strong relationships between [M/H] and colour in $r-z$, $J-K_S$, and identify $W1-W2$ as a metal-sensitive colour over this range. 

Comparison to stellar isochrones shows a lack of agreement in most bands, with the poorest agreement in $u-g$ and $g-r$ and better agreement in $r-z$, $J-K_S$, and $W1-W2$. The \citet{Bochanski2013} empirical $\delta(g-r)$ [M/H] relation is a poor match with the APOGEE KM data, likely because $\delta(g-r)$ was calibrated on subdwarfs, which extend to much lower metallicities than this sample. The \citet{Newton2014} [M/H] relation in $J-K_S$/$H-K_S$ space is a better match to APOGEE KM data, but shows an constant shift between lines of the same metallicity. We present the first $T_{\rm eff}$ and [M/H] relationships based on a combination $r-z$ and $W1-W2$ colours. These initial relations yield $T_{\rm eff}$ to $\sim$100~K and [M/H] to $\sim$0.18~dex precision with colours alone, for $T_{\rm eff}$ in the range of 3550--4200~K and [M/H] in the range of $-$0.5--0.2, and will be substantially improved by refined ASPCAP parameters and an extension to both lower and higher $T_{\rm eff}$ stars. 

\section*{Acknowledgements}
We thank the anonymous referee whose insightful comments have improved this paper. JAJ also acknowledges support from NSF grant AST-0807997.

This work makes use of data from SDSS-III. Funding for SDSS-III has been provided by the Alfred P. Sloan Foundation, the Participating Institutions, the National Science Foundation, and the U.S. Department of Energy Office of Science. The SDSS-III web site is http://www.sdss3.org/.

SDSS-III is managed by the Astrophysical Research Consortium for the Participating Institutions of the SDSS-III Collaboration including the University of Arizona, the Brazilian Participation Group, Brookhaven National Laboratory, Carnegie Mellon University, University of Florida, the French Participation Group, the German Participation Group, Harvard University, the Instituto de Astrofisica de Canarias, the Michigan State/Notre Dame/JINA Participation Group, Johns Hopkins University, Lawrence Berkeley National Laboratory, Max Planck Institute for Astrophysics, Max Planck Institute for Extraterrestrial Physics, New Mexico State University, New York University, Ohio State University, Pennsylvania State University, University of Portsmouth, Princeton University, the Spanish Participation Group, University of Tokyo, University of Utah, Vanderbilt University, University of Virginia, University of Washington, and Yale University.

This publication also makes use of data products from the Two Micron All Sky Survey, which is a joint project of the University of Massachusetts and the Infrared Processing and Analysis Center/California Institute of Technology, funded by the National Aeronautics and Space Administration and the National Science Foundation. 

Additionally, this publication makes use of data products from the Wide-field Infrared Survey Explorer, which is a joint project of the University of California, Los Angeles, and the Jet Propulsion Laboratory/California Institute of Technology, funded by the National Aeronautics and Space Administration.





\bsp	
\label{lastpage}
\end{document}

%% file: apogeeKM_short_v4.tex
\begin{landscape}
\begin{table}
\tiny
\centering
\caption{Properties of APOGEE KM dwarfs} \label{tab:all}
\begin{tabular}{llrlllllllllllll}
\hline
2MASS ID & \multicolumn{2}{c}{ASPCAP} & \multicolumn{5}{c}{SDSS} & \multicolumn{3}{c}{2MASS} & \multicolumn{2}{c}{WISE} & \multicolumn{1}{c}{APASS}  & $V-J^{\rm a}$  & $V-K_S^{\rm b}$ \\
 2M+ & T$_{\rm eff}$ (K) & [M/H] & $u$ & $g$ & $r$ & $i$ & $z$ & $J$ & $H$ & $K_S$ & $W1$ & $W2$ & $V$ & T$_{\rm eff}$ (K) & T$_{\rm eff}$ (K) \\
\hline
00012151+5634379 &  3934  &  0.06  &  \nodata & \nodata & \nodata & \nodata & \nodata & 10.43$\pm$0.02 & 9.72$\pm$0.03 & 9.52$\pm$0.02 & 9.45$\pm$0.02 & 9.48$\pm$0.02 & 13.41 & 3812 & 3768   \\  
00012252+1558339 &  3775  &  $-$0.18  &  19.31$\pm$0.03 & 16.72$\pm$0.02 & 15.29$\pm$0.02 & 14.45$\pm$0.01 & 13.98$\pm$0.02 & 12.76$\pm$0.02 & 12.09$\pm$0.03 & 11.92$\pm$0.02 & \nodata & 11.71$\pm$0.02 & \nodata & \nodata & \nodata   \\  
00012694+1639052 &  3712  &  $-$0.51  &  18.80$\pm$0.03 & \nodata & 15.04$\pm$0.02 & \nodata & 13.71$\pm$0.02 & 12.51$\pm$0.02 & \nodata & \nodata & 11.61$\pm$0.02 & 11.51$\pm$0.02 & 15.58 & 3657 & \nodata   \\  
00013219+0016012 &  4012  &  $-$0.42  &  \nodata & 15.60$\pm$0.01 & \nodata & \nodata & 13.44$\pm$0.01 & \nodata & 11.62$\pm$0.02 & 11.52$\pm$0.02 & 11.39$\pm$0.02 & 11.40$\pm$0.02 & 14.90 & \nodata & 3939   \\  
00013817+0017293 &  3671  &  $-$0.19  &  18.84$\pm$0.03 & 16.32$\pm$0.02 & 14.97$\pm$0.02 & \nodata & 13.28$\pm$0.01 & 12.01$\pm$0.02 & 11.33$\pm$0.02 & 11.10$\pm$0.02 & 11.01$\pm$0.02 & 10.94$\pm$0.02 & 15.62 & 3476 & 3423   \\  
00015592+0027057 &  4046  &  $-$0.11  &  18.99$\pm$0.03 & 16.38$\pm$0.01 & \nodata & 14.44$\pm$0.01 & 14.09$\pm$0.01 & 12.93$\pm$0.02 & 12.23$\pm$0.02 & 12.09$\pm$0.02 & 12.02$\pm$0.02 & 12.05$\pm$0.02 & 15.72 & 3889 & 3850   \\  
00015966+1627449 &  4132  &  $-$0.12  &  \nodata & \nodata & \nodata & \nodata & \nodata & 10.73$\pm$0.02 & \nodata & \nodata & 9.79$\pm$0.02 & 9.85$\pm$0.02 & 13.23 & 4080 & \nodata   \\  
00022557+0126203 &  4068  &  0.06  &  19.45$\pm$0.03 & 16.73$\pm$0.02 & 15.38$\pm$0.01 & 14.68$\pm$0.02 & 14.30$\pm$0.02 & \nodata & 12.50$\pm$0.02 & 12.32$\pm$0.02 & \nodata & 12.25$\pm$0.02 & 15.85 & \nodata & 3953   \\  
00023747-0010572 &  3882  &  $-$0.32  &  19.77$\pm$0.03 & 17.12$\pm$0.02 & 15.72$\pm$0.01 & 15.00$\pm$0.02 & 14.66$\pm$0.02 & 13.40$\pm$0.02 & 12.73$\pm$0.02 & 12.57$\pm$0.02 & \nodata & \nodata & \nodata & \nodata & \nodata   \\  
00025988+0148410 &  4053  &  $-$0.28  &  18.98$\pm$0.03 & 16.34$\pm$0.03 & \nodata & \nodata & \nodata & 12.90$\pm$0.02 & 12.29$\pm$0.02 & \nodata & 12.03$\pm$0.02 & 12.03$\pm$0.02 & 15.51 & 3975 & \nodata   \\  
00030930+0110025 &  3996  &  $-$0.36  &  19.23$\pm$0.03 & 16.65$\pm$0.02 & 15.32$\pm$0.01 & 14.75$\pm$0.01 & 14.44$\pm$0.01 & 13.25$\pm$0.02 & 12.61$\pm$0.03 & 12.43$\pm$0.02 & 12.36$\pm$0.02 & \nodata & \nodata & \nodata & \nodata   \\  
00031412+0037379 &  3930  &  0.04  &  19.10$\pm$0.03 & 16.42$\pm$0.02 & \nodata & 14.60$\pm$0.01 & 14.29$\pm$0.02 & \nodata & \nodata & 12.34$\pm$0.02 & \nodata & 12.27$\pm$0.02 & 15.70 & \nodata & 4045   \\  
00031777+1636147 &  3909  &  $-$0.36  &  18.84$\pm$0.03 & \nodata & 14.95$\pm$0.01 & 14.28$\pm$0.01 & 13.86$\pm$0.01 & 12.68$\pm$0.02 & \nodata & 11.86$\pm$0.02 & \nodata & 11.75$\pm$0.02 & 15.54 & 3801 & 3764   \\  
00033020+0020078 &  3964  &  $-$0.10  &  19.55$\pm$0.03 & 16.89$\pm$0.02 & \nodata & \nodata & 14.53$\pm$0.02 & 13.33$\pm$0.02 & 12.69$\pm$0.03 & 12.44$\pm$0.02 & 12.41$\pm$0.02 & 12.45$\pm$0.02 & \nodata & \nodata & \nodata   \\  
00033817+0020226 &  3907  &  $-$0.18  &  19.14$\pm$0.02 & 16.49$\pm$0.02 & \nodata & \nodata & 14.00$\pm$0.02 & 12.81$\pm$0.02 & 12.13$\pm$0.03 & 11.94$\pm$0.02 & 11.86$\pm$0.02 & 11.85$\pm$0.02 & 15.78 & 3772 & 3717   \\  
00035823+7351001 &  4114  &  $-$0.17  &  \nodata & \nodata & \nodata & \nodata & \nodata & \nodata & \nodata & 11.96$\pm$0.02 & \nodata & 11.87$\pm$0.02 & 16.08 & \nodata & 3586   \\  
00035968+1542051 &  3924  &  $-$0.26  &  18.86$\pm$0.03 & 16.34$\pm$0.02 & \nodata & 14.34$\pm$0.02 & 13.99$\pm$0.02 & 12.78$\pm$0.02 & 12.09$\pm$0.03 & 11.99$\pm$0.02 & 11.92$\pm$0.02 & 11.91$\pm$0.02 & 15.62 & 3830 & 3809   \\  
00041959+7547098 &  3545  &  $-$0.30  &  \nodata & \nodata & \nodata & \nodata & \nodata & 12.11$\pm$0.02 & 11.54$\pm$0.03 & 11.30$\pm$0.02 & 11.15$\pm$0.02 & 11.00$\pm$0.02 & \nodata & \nodata & \nodata   \\  
00042083+0158446 &  4133  &  $-$0.23  &  \nodata & \nodata & \nodata & \nodata & \nodata & \nodata & 11.96$\pm$0.03 & 11.78$\pm$0.02 & 11.70$\pm$0.02 & 11.73$\pm$0.02 & 15.08 & \nodata & 4026   \\  
00043956+1525247 &  3550  &  $-$0.29  &  18.16$\pm$0.02 & 15.72$\pm$0.01 & \nodata & \nodata & 12.64$\pm$0.02 & 11.30$\pm$0.02 & 10.77$\pm$0.03 & 10.51$\pm$0.02 & 10.39$\pm$0.02 & 10.31$\pm$0.02 & 14.98 & 3432 & 3410   \\  
00044424+0038241 &  3786  &  $-$0.04  &  19.41$\pm$0.03 & 16.84$\pm$0.02 & 15.46$\pm$0.01 & 14.78$\pm$0.02 & 14.41$\pm$0.02 & 13.23$\pm$0.02 & 12.59$\pm$0.02 & 12.36$\pm$0.02 & \nodata & \nodata & \nodata & \nodata & \nodata   \\  
00044471-0011336 &  4022  &  $-$0.13  &  18.56$\pm$0.02 & 15.90$\pm$0.02 & \nodata & \nodata & 13.49$\pm$0.01 & 12.33$\pm$0.02 & 11.63$\pm$0.03 & 11.48$\pm$0.02 & 11.39$\pm$0.02 & 11.42$\pm$0.02 & 15.10 & 3901 & 3849   \\  
00044671+0125326 &  4005  &  $-$0.27  &  \nodata & \nodata & \nodata & \nodata & \nodata & 12.93$\pm$0.02 & 12.27$\pm$0.02 & 12.10$\pm$0.02 & \nodata & 12.00$\pm$0.02 & 15.52 & 3985 & 3942   \\  
00044884-0032341 &  4073  &  $-$0.29  &  \nodata & 16.49$\pm$0.02 & 15.16$\pm$0.02 & 14.58$\pm$0.02 & 14.26$\pm$0.01 & 13.13$\pm$0.02 & 12.48$\pm$0.03 & \nodata & \nodata & 12.22$\pm$0.02 & 15.71 & 3993 & \nodata   \\  
00054076+0001181 &  4024  &  $-$0.37  &  18.33$\pm$0.02 & 15.73$\pm$0.01 & \nodata & \nodata & 13.45$\pm$0.02 & 12.25$\pm$0.02 & 11.60$\pm$0.02 & 11.45$\pm$0.02 & \nodata & 11.37$\pm$0.02 & 14.96 & 3897 & 3873   \\  
00054249+0022537 &  4003  &  $-$0.25  &  19.34$\pm$0.03 & 16.78$\pm$0.02 & 15.45$\pm$0.02 & \nodata & 14.49$\pm$0.02 & \nodata & 12.64$\pm$0.02 & 12.48$\pm$0.02 & 12.45$\pm$0.02 & \nodata & \nodata & \nodata & \nodata   \\  
00055969-0030062 &  4027  &  $-$0.38  &  19.15$\pm$0.03 & 16.51$\pm$0.03 & 15.15$\pm$0.02 & 14.60$\pm$0.01 & \nodata & 13.10$\pm$0.02 & 12.45$\pm$0.02 & 12.28$\pm$0.02 & \nodata & 12.19$\pm$0.02 & \nodata & \nodata & \nodata   \\  
00060369+0104479 &  4105  &  $-$0.48  &  \nodata & \nodata & \nodata & \nodata & \nodata & 12.90$\pm$0.02 & 12.23$\pm$0.03 & 12.05$\pm$0.02 & 11.98$\pm$0.02 & 12.00$\pm$0.02 & \nodata & \nodata & \nodata   \\  
00060971+0120321 &  3745  &  0.09  &  \nodata & \nodata & \nodata & \nodata & \nodata & \nodata & 12.67$\pm$0.03 & \nodata & 12.32$\pm$0.02 & \nodata & \nodata & \nodata & \nodata   \\  
00061237+0101599 &  3968  &  $-$0.16  &  19.29$\pm$0.03 & 16.68$\pm$0.02 & 15.29$\pm$0.02 & 14.61$\pm$0.01 & 14.21$\pm$0.01 & 13.03$\pm$0.02 & 12.32$\pm$0.03 & \nodata & 12.07$\pm$0.02 & 12.07$\pm$0.02 & 15.80 & 3893 & \nodata   \\  
\hline
\multicolumn{16}{l}{$^{\rm a}$ from \citet{Boyajian2013}} \\
\multicolumn{16}{l}{$^{\rm b}$ from \citet{Mann2015}}
\end{tabular}
This table is a shortened version provided as a guide. Complete table available online.
\end{table}
\end{landscape}